\documentclass[12pt]{article}
 \usepackage{epsfig}
 \def\beq{\begin{equation}}
 \def\eeq{\end{equation}}
 \def\bea{\begin{eqnarray}}
 \def\eea{\end{eqnarray}}
 \def\nn{\nonumber}
 \def\sss{\scriptscriptstyle}
 
 \def\bd{B_d^0}
 \def\bdbar{{\bar B}_d^0}

 \def\barp{{\raise.35ex\hbox
 {${\sss (}$}}---{\raise.35ex\hbox{${\sss )}$}}}
 \def\bdbarp{\hbox{$B_d$\kern-1.4em\raise1.4ex\hbox{\barp}}}
 \def\bsbarp{\hbox{$B_s$\kern-1.4em\raise1.4ex\hbox{\barp}}}
 \def\ks{K_{\sss S}}
 
 \def\barpk{{\raise.35ex\hbox
 {${\sss (}$}}--{\raise.35ex\hbox{${\sss )}$}}}
 \def\kbarp{\hbox{$K$\kern-0.9em\raise1.4ex\hbox{\barpk}}}
 \def\roughly#1{\mathrel{\raise.3ex\hbox
 {$#1$\kern-.75em\lower1ex\hbox{$\sim$}}}}

 \def\adir00{{a_{\sss dir}^{00}}}

 \def\B00{B^{00}}
 \def\Bp0{B^{+0}}

 \def\epjc#1#2#3{{\it Eur.\ Phys.\ J.}\ {\bf C#1} (19#2) #3}

 \def\plb#1#2#3{{\it Phys.\ Lett.} {\bf #1B} (19#2) #3}

 \def\prd#1#2#3{{\it Phys.\ Rev.} {\bf D#1} (19#2) #3}

 \def\prl#1#2#3{{\it Phys.\ Rev.\ Lett.} {\bf #1} (19#2) #3}

\def\beq{\begin{equation}}
\def\eeq#1{\label{#1}\end{equation}}
\def\eeqn{\end{equation}}

\def\beqa{\begin{eqnarray}}
\def\eeqa#1{\label{#1}\end{eqnarray}}
\def\eeqan{\end{eqnarray}}

\let\bar=\overbar

\def\Dslash{\not{\hbox{\kern-4pt $D$}}}
\def\dslash{\not{\hbox{\kern-2pt $\del$}}}

\def\msb{{\bar{\ssstyle M \kern -1pt S}}}

\def\Title#1{\begin{center} {\Large {\bf #1} } \end{center}}

\begin{document}

\begin{flushright}
UdeM-GPP-TH-02-99 \\
IMSc-2002/06/14 \\
\end{flushright}

\Title{New Physics in $B\to J/\psi K^*$\footnote{Talk given by Rahul
Sinha at {\it Flavor Physics and CP Violation (FPCP)}, Philadelphia,
PA, USA, May 2002}}


\bigskip\bigskip


\begin{raggedright}  

{\it David London\index{Sinha, Rahul}\\
Laboratoire Ren\'e J.-A. L\'evesque \\
Universit\'e de Montr\'eal \\
C.P. 6128, succ.\ centre-ville \\
Montr\'eal, QC, CANADA H3C 3J7\\
\bigskip
Nita Sinha and Rahul Sinha\\
The Institute of Mathematical Sciences\\
Taramani, Chennai 600113 INDIA}
\bigskip\bigskip
\end{raggedright}


CP violation in the $B$ system \cite{CPreview} has been established by
the recent measurements of $\sin 2\beta$, with a world average of
$\sin 2\beta=0.78\pm 0.08$ \cite{betameas}. The main $B$ decay used to
probe the weak phase $\beta$ is the so-called ``gold-plated'' mode
$\bd(t)\to J/\psi \ks$.  In order to extract weak-phase information
cleanly, i.e.\ with no hadronic uncertainties, a given $B$ decay must
be dominated by a single weak decay amplitude. However, even within
the SM, $\bd\to J/\psi \ks$ receives contributions from two weak
amplitudes: the tree amplitude and the $b\to s$ penguin amplitude.
Nevertheless, this decay mode is very clean for the following two
reasons.  First, the $c{\bar c}$ quark pair must be produced in a
color-singlet state, requiring three gluons in the penguin
amplitude. Consequently, the penguin contribution is expected to be
considerably smaller than the tree contribution. Second, in the
Wolfenstein parameterization \cite{Wolfenstein} the CKM matrix elements involved in the
$b\to s$ penguin amplitude ($V_{tb}^* V_{ts}$) and in the tree
amplitude ($V_{cb}^* V_{cs}$) are both real. Thus, the weak phases of
these two amplitudes are the same, so that effectively only a single
weak amplitude contributes to $\bd\to J/\psi \ks$. The extraction of
the CP phase $\beta$ from this decay mode is therefore extremely
clean.

The decay $\bd(t) \to J/\psi K^*$ is also a clean mode, for exactly
the same reasons as above. The complication, in comparison to
$\bd(t)\to\psi\ks$, is that the final state now consists of two vector
particles, so that the CP-even and CP-odd components must be
distinguished by performing an angular analysis \cite{CPreview}. Each
component can then be treated separately, and $\beta$ can be obtained
cleanly.

In the presence of new physics, the extraction of the weak phase
$\beta$ may not be clean.  If the new physics contributes only to
$\bd$--$\bdbar$ mixing, the measurement of $\beta$ remains clean,
though the measured value is not the true SM value, but rather one
which is shifted by a new-physics phase. On the other hand, if new
physics affects the decay amplitude, then the extraction of $\beta$ is
no longer clean -- it may be contaminated by hadronic
uncertainties. It is this situation which interests us.

How can new physics affect the decay amplitude? This can occur if
there are new contributions to the $b\to s$ penguin
amplitude\cite{newphysics}, so that this amplitude no longer has the
same weak phase as the tree amplitude. There are a variety of
new-physics models in which this can occur. These include, for
example, supersymmetric models with R-parity breaking, $Z$- and
$Z'$-mediated flavor-changing neutral currents
\cite{LerLon}, and the Top-Higgs doublet model \cite{Kuo-das}.

An obvious question is then: how does one see new-physics
contributions to the decay amplitudes if they are present? The
standard method is to search for direct CP violation. In the presence
of two decay amplitudes, the full amplitude for the decay $B \to f$
can be written as
\begin{equation} A(B \to f) = a e^{i \phi_a} e^{i\delta^a} + b e^{i \phi_b}
e^{i\delta^b} ~.  
\end{equation} 
Here, $\phi_{a,b}$ and $\delta^{a,b}$ are, respectively, the weak and
strong phases of the two contributing amplitudes. The amplitude for
the CP-conjugate decay $\bar{B} \to {\bar f}$ can be obtained from the
above by changing the signs of the weak phases. The direct CP
asymmetry  $a^{CP}_{dir}$ is then given by
\begin{equation}
 a^{CP}_{dir}= {\Gamma(B \to f) - \Gamma({\bar B} \to
{\bar f}) \over \Gamma(B \to f) + \Gamma({\bar B} \to {\bar f})} = - {2
a b \sin(\phi_a-\phi_b) \sin(\delta^a-\delta^b) \over a^2 +
b^2 + 2 a b \cos(\phi_a-\phi_b) \cos(\delta^a-\delta^b)} ~.
 \label{directcp}
\end{equation}
This expression holds for both neutral and charged $B$ decays. Thus,
if new physics is present, we can expect to see direct CP violation in
both $\bdbarp \to J/\psi \ks$ and $B^\pm \to J/\psi K^\pm$ decays.

It is obvious from Eq.~(\ref{directcp}) that an observable direct
asymmetry requires not only a nonzero weak-phase difference between
the two decay amplitudes, but also a strong-phase difference. However,
it has been argued that since the $b$-quark is rather heavy, all
strong phases in $B$ decays should be quite small. If the strong
phases of the two amplitudes happen to be almost equal, there will be
no observable signal of direct CP violation, even though new physics
is present. Hence new physics may be hard to find using direct
asymmetries.

One is therefore led to the question: {\it if the strong-phase
differences vanish, is there any way of detecting the presence of new
physics?} As we show below, the answer to this question is {\it yes},
if one uses the final state $J/\psi K^*$ rather than $J/\psi K$. As
above, we assume that there are two contributions to the decay
amplitude, coming from the SM and from new physics. The weak
phase of the SM contribution is zero, while the new physics contribution has a
nonzero weak phase. Since the final state consists of two vector
mesons, there are three helicity amplitudes. These take the form
\bea 
A_\lambda \equiv Amp (B \to J/\psi K^*)_\lambda &=& a_\lambda
 e^{i \delta_\lambda^a} + b_\lambda e^{i\phi} e^{i \delta_\lambda^b}
 ~, \nn\\ {\bar A}_\lambda \equiv Amp ({\bar B} \to J/\psi {\bar
 K}^*)_\lambda &=& a_\lambda e^{i \delta_\lambda^a} + b_\lambda
 e^{-i\phi} e^{i \delta_\lambda^b} ~,
 \label{amps}
\eea
where the $a_\lambda$ and $b_\lambda$ represent the SM and new physics
amplitudes, $\phi$ is the new-physics weak phase, the
$\delta_\lambda^{a,b}$ are the strong phases, and the helicity index
$\lambda$ takes the values $\left\{ 0,\|,\perp \right\}$. Using CPT
invariance, the full decay amplitudes can be written as
\bea
 {\cal A} = Amp (B\to J/\psi K^*) = A_0 g_0 + A_\| g_\| + i \, A_\perp
 g_\perp~~ ,\\
 {\bar{\cal A}} = Amp ({\bar B} \to J/\psi {\bar K}^*) = {\bar A}_0 g_0
      + {\bar A}_\| g_\| - i \, {\bar A}_\perp g_\perp~~ ,
 \label{fullamps}
\eea
where the $g_\lambda$ are the coefficients of the helicity amplitudes
written in the linear polarization basis. The $g_\lambda$ depend only
on the angles describing the kinematics \cite{flambda,flamdun}.

We first consider neutral $B$ decays and assume that the $\kbarp^*$ is
detected through its decay to $\ks \pi^0$, so that both $\bd$ and
$\bdbar$ decay to the same final state. With the above equations, the
time-dependent decay rates for $\bdbarp(t) \to J/\psi \kbarp^*$ can be
written as
\begin{equation}
 \Gamma(\bdbarp(t) \to J/\psi K^*) =  e^{-\Gamma t}
 \sum_{\lambda\leq\sigma}
 \Bigl(\Lambda_{\lambda\sigma} \pm \Sigma_{\lambda\sigma}\cos(\Delta M t) 
\mp \rho_{\lambda\sigma}\sin(\Delta M t)\Bigr) g_\lambda g_\sigma ~.
 \label{decayrates}
\end{equation}
 By performing a time-dependent study and angular analysis of the
decays $\bdbarp(t)\to J/\psi \kbarp^*$, one can measure the
observables $\Lambda_{\lambda\sigma}$, $\Sigma_{\lambda\sigma}$ and
$\rho_{\lambda\sigma}$. In terms of the helicity amplitudes
$A_0,A_\|,A_\perp$, these can be expressed as follows:
\bea
 &\Lambda_{\lambda\lambda}=\displaystyle
 \frac{|A_\lambda|^2+|{\bar A}_\lambda|^2}{2}~,&
 \Sigma_{\lambda\lambda}=\displaystyle
 \frac{|A_\lambda|^2-|{\bar A}_\lambda|^2}{2}~,\nn \\[1.5ex]
 &\Lambda_{\perp i}= -\!{\rm Im}({ A}_\perp { A}_i^* \!-\! {\bar A}_\perp
 {{\bar A}_i}^* )~,
 &\Lambda_{\| 0}= {\rm Re}(A_\| A_0^*\! +\! {\bar A}_\| {{\bar A}_0}^* )~,
 \nn \\[1.5ex]
 &\Sigma_{\perp i}= -\!{\rm Im}(A_\perp A_i^*\! +\! {\bar A}_\perp {{\bar
 A}_i}^* )~,
 &\Sigma_{\| 0}= {\rm Re}(A_\| A_0^*\!-\! {\bar A}_\| {{\bar A}_0}^*
 )~,\nn\\[1.5ex]
 &\rho_{\perp i}\!=\!-\!{\rm Re}\!\Bigl(\!\frac{q}{p}
 \![A_\perp^*  {\bar A}_i\! +\! A_i^* {\bar A}_\perp\!]\!\Bigr)~,
 &\rho_{\perp \perp}\!=\! -\! {\rm Im}\Bigl(\frac{q}{p}\,
 A_\perp^* {\bar A}_\perp\Bigr)~,\nn\\[1.5ex]
 &\rho_{\| 0}\!=\!{\rm Im}\!\Bigl(\frac{q}{p}\!
 [A_\|^* {\bar A}_0\! + \!A_0^* {\bar A}_\|\!]\!\Bigr)~,
 &\rho_{ii}\!=\!{\rm Im}\!\Bigl(\frac{q}{p} A_i^* {\bar A}_i\Bigr)~,
 \label{observables}
\eea
where $i=\{0,\|\}$. In the above, $q/p = \exp({-2\,i\phi_M})$, where
$\phi_M$ is the weak phase in $\bd$--$\bdbar$ mixing (in the SM,
$\phi_M = \beta$). Note that the direct CP asymmetry $a^{CP}_{dir}$ is
proportional to the $\Sigma_{\lambda\lambda}$ observables.

The key observable in the $B\to J/\Psi K^*$ mode is $\Lambda_{\perp
i}$. Using the expressions for the amplitudes found in
Eq.~(\ref{amps}), this can be written as
\begin{equation}
 \Lambda_{\perp i} = 2 \left[ a_\perp b_i \cos(\delta_\perp^a -
 \delta_i^b) - a_i b_\perp \cos(\delta_\perp^b - \delta_i^a) \right]
 \sin\phi ~.
\end{equation}
Even if the strong-phase differences vanish, this observable is still
nonzero in the presence of new physics ($\phi\ne 0$), in contrast to
the direct CP asymmetry $a^{CP}_{dir}$ given in Eq.~(\ref{directcp}).
Thus, a complete search for new physics should include the measurement
of $\Lambda_{\perp i}$ in addition to $a^{CP}_{dir}$.

The reason that $\Lambda_{\perp i}$ is proportional to
$\cos(\delta_\perp^{a,b} - \delta_i^{b,a})$, rather than
$\sin(\delta_\perp^{a,b} - \delta_i^{b,a})$, is that the $\perp$
helicity is CP-odd, while the 0 and $\|$ helicities are CP-even. Thus,
$\perp$--0 and $\perp$--$\|$ interferences are CP-odd and switch sign
\cite{flambda,triple,BD,HH} between process and conjugate process. This results in
the $\cos(\delta_\perp^{a,b} - \delta_i^{b,a})$ term. Obviously, such
interferences will not occur for final states such as $J/\psi \ks$,
which have only one helicity state.

As can be seen from Eq.~(\ref{decayrates}), the $\Lambda_{\perp i}$
term is common to both $\bd(t)$ and $\bdbar(t)$ decay rates. Thus, if
one does not distinguish between $\bd(t)$ and $\bdbar(t)$ decays, and
instead simply adds the two rates together, the
$\Lambda_{\lambda\sigma}$ terms remain. Note also that these terms are
time-independent (i.e.\ they are not proportional to $\cos\Delta Mt$
or $\sin \Delta M t$). Therefore, {\it no tagging or time-dependent
measurements are needed to extract $\Lambda_{\perp i}$}! It is only
necessary to perform an angular analysis of the final state $J/\psi
\kbarp^*$, with $\kbarp^* \to \ks \pi^0$. Thus, this measurement can even be
made at a symmetric $B$-factory such as CLEO.

The decays $B^\pm \to J/\psi K^{*\pm}$ are even simpler to analyze
since no mixing is involved. It is straightforward to see that it is
not even necessary to distinguish between $B^+$ and $B^-$ decays for
this measurement. In light of this, one can in principle combine
charged and neutral $B$ decays to increase the sensitivity to new
physics. One simply performs an angular analysis on all decays in
which a $J/\psi$ is produced accompanied by a charged or neutral
$K^*$. A nonzero value of $\Lambda_{\perp i}$ is a smoking-gun signal
for new physics.
 
Now, suppose that $\Lambda_{\perp i}$ is measured to be nonzero. This
means that new physics is present, which in turn implies that the
measured value of $\beta$ as extracted from $\bd(t)\to J/\psi \ks$ or
$\bd(t) \to \psi K^*$ is not the true SM value of $\beta$. This then
raises the following questions. Is it nevertheless possible to obtain
the true value of $\beta$ from measurements of $\bd(t) \to \psi K^*$?
If not, can one at least constrain the difference $|\beta -
\beta^{meas}|$? We explore these questions below.

It is straightforward to show that one cannot extract the true value
of $\beta$. There are a total of six amplitudes describing $\bd(t) \to
\psi K^*$ [Eq.~(\ref{amps})]. Experimentally, at best one can measure
the magnitudes and relative phases of these six amplitudes, giving 11
measurements. However, there are a total of 13 theoretical parameters
describing these amplitudes: 3 $a_\lambda$'s, 3 $b_\lambda$'s, 5
strong phase differences, $\phi$ and $\beta$. Since there are more
unknown parameters than there are measurements, one cannot obtain any
of the unknowns. In particular, it is impossible to extract $\beta$.

However, it is still possible to constrain the theoretical parameters. 
For example, with a bit of algebra one can express $b_\lambda$ as follows:
\begin{equation}
b_\lambda^2 = {1\over 2\sin^2\phi} \left[ \Lambda_{\lambda\lambda} -
\sqrt{\Lambda_{\lambda\lambda}^2 - \Sigma_{\lambda\lambda}^2}
\cos(2\beta^{meas}_\lambda-2\beta) \right].
\end{equation}
The minimum value of $b_\lambda^2$ is easy to find:
\begin{equation}
b^2_\lambda \ge {1\over 2} \left[ \Lambda_{\lambda\lambda} -
\sqrt{\Lambda_{\lambda\lambda}^2 - \Sigma_{\lambda\lambda}^2} \right].
\end{equation}
Thus, if direct CP violation is observed ($\Sigma_{\lambda\lambda} \ne
0$), one can place a lower bound on the new-physics amplitude
$b_\lambda$, and consequently the scale of new physics.  The above
bound becomes trivial, i.e.\ $b_\lambda \ge 0$, if all strong phases
are quite small, leading to a vanishing value of
$\Sigma_{\lambda\lambda}$.  However, even if the strong phases vanish,
it is still possible to obtain lower bounds on $b_\lambda$ and $|\beta
- \beta^{meas}|$ using measurements of $\Lambda_{\perp i}$
\cite{lss-8}.

\bigskip

R.S. thanks the organizers of FPCP2002 for a wonderful
conference. N.S. and R.S. thank D.L. for the hospitality of the
Universit\'e de Montr\'eal, where part of this work was done. The work
of D.L. was financially supported by NSERC of Canada.  The work of
Nita Sinha was financially supported by a young scientist award of the
Department of Science and Technology, India.

\end{document}